\documentclass[aps,prc,reprint,superscriptaddress,floatfix,showpacs,nobibnotes]{revtex4-1}
\usepackage{graphicx,caption}
\usepackage{graphics}
\usepackage{epsfig}
\usepackage{amssymb}
\usepackage{amsmath}
\usepackage{amsfonts}
\usepackage{hyperref}
\usepackage{textcomp}
\usepackage{subfig}
\usepackage{tabularx}
\usepackage{xtab}
\usepackage{latexsym}

\begin{document}

\title{Isomers in $^{203}$Tl and core excitations built on a five-nucleon-hole structure}

\author{V. Bothe}
\affiliation{School of Physical Sciences, UM-DAE Centre for Excellence in Basic Sciences, Mumbai 400098, India} 
\author{S.K. Tandel}
\email{Electronic address: sujit.tandel@cbs.ac.in ; sktandel@gmail.com}
\affiliation{School of Physical Sciences, UM-DAE Centre for Excellence in Basic Sciences, Mumbai 400098, India} 
\affiliation{Department of Physics, University of Massachusetts Lowell, Lowell, Massachusetts 01854, USA}
\author{S.G. Wahid}
\affiliation{School of Physical Sciences, UM-DAE Centre for Excellence in Basic Sciences, Mumbai 400098, India}
\author{P.C. Srivastava}
\affiliation{Department of Physics, Indian Institute of Technology Roorkee, Roorkee 247667, India}
\author{Bharti Bhoy}
\affiliation{Department of Physics, Indian Institute of Technology Roorkee, Roorkee 247667, India}
\author{P. Chowdhury} 
\affiliation{Department of Physics, University of Massachusetts Lowell, Lowell, Massachusetts 01854, USA}
\author{R.V.F. Janssens}
\affiliation{Department of Physics and Astronomy, University of North Carolina at Chapel Hill, Chapel Hill, North Carolina 27599, USA}
\affiliation{Triangle Universities Nuclear Laboratory, Duke University, Durham, North Carolina 27708, USA}
\author{F.G. Kondev}
\affiliation{Argonne National Laboratory, Argonne, Illinois 60439, USA}
\author{M.P. Carpenter} 
\affiliation{Argonne National Laboratory, Argonne, Illinois 60439, USA}
\author{T. Lauritsen}
\affiliation{Argonne National Laboratory, Argonne, Illinois 60439, USA}
\author{D. Seweryniak}
\affiliation{Argonne National Laboratory, Argonne, Illinois 60439, USA}
\author{S. Zhu}
\affiliation{Argonne National Laboratory, Argonne, Illinois 60439, USA}
\date{\today}

\vspace{0.5in}

\begin{abstract}
Isomers with three- and five-nucleon-hole configurations have been established in $^{203}$Tl. 
These include newly identified levels with a three-nucleon structure: {\it I}$^{\pi }$ = (15/2$^{-}$) 
with {\it T}$_{1/2}$ = 7.9(5) ns, and {\it I}$^{\pi }$ = (35/2$^{-}$) with {\it T}$_{1/2}$ = 4.0(5) ns.
In addition, five-quasiparticle states: {\it I}$^{\pi }$ = (39/2$^{-}$) with {\it T}$_{1/2}$ = 1.9(2) ns, 
and {\it I}$^{\pi }$ = (49/2$^{+}$) with {\it T}$_{1/2}$ = 3.4(4) ns have also been established.
The previously determined long-lived decay [{\it T}$_{1/2}$ = 6.6(3) $\mu $s from this work] 
is associated with isomerism of the {\it I}$^{\pi }$ = (29/2$^{+}$) state. Levels above this
long-lived isomer have been identified through a delayed-prompt coincidence measurement. 
Five-nucleon-hole states with excitation energies {\it E}$_{x}$ $\approx $ 7 MeV have been established 
as well as possible octupole excitations of the $^{208}$Pb core built on these levels. 
The level scheme of $^{203}$Tl is extended up to {\it E}$_{x}$ $\approx $ 11 MeV with the 
inclusion of 25 new transitions. Empirical and shell-model calculations have been performed to aid in the 
description of the observed states which are found to be predominantly of intrinsic character.
\end{abstract}


\maketitle

\section{Introduction}
The occurrence of metastable states in nuclei with proton or neutron numbers in the 
vicinity of shell closures is well established as arising from the presence of adjacent 
excited levels with a relatively large difference in angular momentum and small 
difference in energy. Isotopes of Hg ({\it Z} = 80) and
Tl ({\it Z} = 81) with {\it A} $\approx $ 200, lying just below the doubly-magic
$^{208}$Pb nucleus in the periodic chart, are therefore good candidates for the observation
of metastable states or isomers \cite{Hausser,Wrzesinski1,Broda,Szpak,Wrzesinski2}.
The proton-rich isotopes of Tl ({\it A} $<$ 200) are characterized by moderate oblate 
deformation on account of a significant number of valence neutrons leading to a 
small degree of collectivity \cite{Kreiner1,Lawrie}. However, when approaching the
{\it N} = 126 shell gap, collective behavior is no longer evident. This is the case
for isotopes of Tl near the line of stability, where $^{203}$Tl and $^{205}$Tl are the 
two stable isotopes \cite{Hausser,Wrzesinski1,Broda,Wahid1,Slocombe}.
The nucleus $^{203}$Tl, with {\it N} = 122, is only one proton and four neutron holes 
away from doubly-magic $^{208}$Pb and, as a result, its structure is expected to 
be dominated by single-particle excitations. The available valence orbitals are the 
proton {\it s}$_{1/2}$, {\it d}$_{3/2}$, {\it d}$_{5/2}$ and {\it h}$_{11/2}$ states
and the neutron {\it p}$_{1/2}$, {\it p}$_{3/2}$, {\it f}$_{5/2}$ and 
{\it i}$_{13/2}$ ones. The presence of the unique-parity $\pi ${\it h}$_{11/2}$
and $\nu ${\it i}$_{13/2}$ orbitals allows for the realization of yrast,
high-spin states, some of which may be of isomeric character, owing to hindrance
induced by the angular momentum selection rule or a change in configuration. 

The systematic delineation of isomers along an
isotopic chain, most of which have relatively pure intrinsic character, provides a
host of nuclear structure information, among which figure residual interactions between 
nucleons occupying different subshells. 
Heavy-ion and $\alpha $-induced reactions have been used to populate excited states
at high spin in proton-rich Tl isotopes \cite{Lawrie,Kreiner2}. However,
in order to study isotopes near the line of stability, different reaction mechanisms
such as inelastic excitation and multi-nucleon transfer may be utilized. In order to 
populate states with high spin, heavy projectiles with above-barrier energies are required.
The isotopes of Tl approaching the shell closure at {\it N} = 126,
particularly $^{204,205}$Tl \cite{Broda,Wrzesinski1}, have been studied up to high 
spin and several metastable states established. In contrast, there was limited 
knowledge on isomers in $^{200-203}$Tl prior to investigations by this collaboration. 
Recently, the observation of isomers with half-lives ranging from a few nanoseconds to 
hundreds of microseconds in $^{200}$Tl \cite{Roy} and $^{202}$Tl \cite{Wahid1}
were reported. The existing information on $^{203}$Tl prior to
the present work is limited, with levels up to possible spin (17/2) $\hbar $ and excitation 
energy $\approx $3 MeV established through a deuteron-induced reaction \cite{Slocombe},
while a microsecond isomer was identified using projectile fragmentation \cite{Pfutzner}.
In the former instance, the presence of a nanosecond isomer has been inferred,
but its location was not established. In the latter case, no level scheme 
is reported, and a tentative assignment of {\it I}$^{\pi }$ = (25/2$^{+}$) and
a three-nucleon-hole character for the microsecond isomer is proposed.
The motivation for the present work was to obtain unambiguous information on
the structural aspects mentioned above in order to advance the understanding of the
structure of $^{203}$Tl. An additional goal was to extend the level scheme of
$^{203}$Tl to also encompass levels arising from the maximum possible number of
nucleon-hole excitations, which is five in this case, which allows for discriminating
tests of interactions used in shell-model calculations. Furthermore, it was also proposed 
to investigate the presence of possible octupole core excitations in $^{203}$Tl
which would be built on five-nucleon-hole configurations. In this region, such excitations
have thus far been observed only in $^{203}$Hg \cite{Szpak}.

\section{Experiment and Data Analysis}
Multi-nucleon transfer reactions were used to populate excited states in $^{203}$Tl, with a
1450-MeV $^{209}$Bi beam from the ATLAS accelerator at Argonne National Laboratory, incident 
on a thick, 50 mg/cm$^2$ Au target. Gamma-ray coincidence data were recorded with the
Gammasphere array comprising of 100 Compton-suppressed high-purity germanium detectors 
\cite{Lee,Janssens}. In addition to prompt coincidence events, delayed data from the decay
of isomeric levels were also recorded. In the first instance, the $^{209}$Bi beam, with the 
natural 82.5 {\it ns} pulsing from the ATLAS superconducting linear accelerator, 
was incident on the target for $\approx $1 {\it ns}, 
and then swept away for a 825-{\it ns} off period. Three- and higher-fold coincidence 
events were recorded within a 1 $\mu $s coincidence window, allowing for the measurement of 
half-lives ranging from a few to several hundred nanoseconds. Later, two- and higher-fold
coincidence data were collected only during beam-off periods by sweeping the beam with
an ON/OFF ratio of 200 $\mu $s/800 $\mu $s via a free-running external clock, enabling the 
study of isomers with long ($\mu $s) half-lives.

The coincidence data were sorted into a variety of histograms in order to establish the level
structure and determine the half-lives of isomers using the TSCAN \cite{Jin} and
Radware \cite{Radford} suite of programs. Three-dimensional histograms of prompt and 
delayed coincidence events were created and analyzed, corresponding to detection of all 
three $\gamma $ rays within $\pm $40 ns and 50-650 ns of the trigger, respectively. To 
identify $\gamma $ rays feeding the $\mu $s isomer, a three-dimensional histogram requiring 
two delayed transitions together with a prompt one, was constructed. The centroid-shift 
method was used to determine half-lives ranging between $\approx $1-10 ns 
using an energy-gated two-dimensional 
histogram with energy on one axis and time difference of the associated transitions 
on the other. The coincidence relationships between $\gamma $ rays deexciting 
the $\mu $s isomer were also analyzed with the help of a three-dimensional energy 
histogram consisting of transitions detected in the 
800 $\mu $s beam-off period. To determine the half-life of the $\mu $s isomer, several 
three-dimensional histograms spanning suitable time intervals within the 800 $\mu $s 
beam-off period were created. Further details of data analysis methods are described
elsewhere \cite{Tandel1,Tandel2,Wahid2}.

\section{Results}
Excited states in $^{203}$Tl up to medium spin and excitation energy of 
$E_x$ $\approx $ 3 MeV had been identified earlier through 
the $^{204}$Hg(d,3n) reaction with an incident deuteron energy of $\approx $ 25 MeV 
\cite{Slocombe}, and projectile fragmentation using a 1 GeV/nucleon $^{238}$U beam 
on a thick $^{9}$Be target \cite{Pfutzner}. In the former, a cascade of
$\gamma $ rays with transition energies of 588, 533, 328, 350 and 265 keV were placed,
in that order, above the 11/2$^{-}$ level at 1450 keV. The presence of an isomer with 
{\it T}$_{1/2}$ = 7.9$^{+10}_{-13}$ ns had been inferred, however its precise
location had not been determined. From the projectile fragmentation work, feeding from
a long-lived, {\it T}$_{1/2}$ = 7.7(5) $\mu $s isomer was reported and a spin-parity,
{\it I}$^{\pi }$ = (25/2$^{+}$), was suggested. Though the above $\gamma $ rays were 
observed, a level scheme was not reported.

The level scheme for $^{203}$Tl deduced from the present work is shown in Fig. 1. 
The presence of the two previously reported isomers is confirmed,
and a slightly improved value of half-life is deduced for the $\mu $s isomer. The placement
of the previously reported $\gamma $ rays and level energies above the 588-keV, 
13/2$^{-}$ $\rightarrow $ 11/2$^{-}$ transition, have been revised as described below, 
based on observed $\gamma $ rays and their coincidence relationships in the
three-dimensional energy histogram. A new level with $E_{x}$ = 2048 keV and possible
spin-parity {\it I}$^{\pi }$ = (15/2$^{-}$) is established. Several decay branches are
observed from this state, including the 598-, 486- and 477-keV $\gamma $ rays, along with 
an unobserved 10-keV transition to the {\it I}$^{\pi }$ = 13/2$^{-}$ level. 
From previous work \cite{Slocombe,Pfutzner}, 
the 350- and 265-keV $\gamma $ rays were inferred to be in
cascade above the 328- and 533-keV transitions. The present data unambiguously indicate 
that these 350- and 265-keV $\gamma $ rays are not in mutual coincidence 
although they are coincident with all lower- and higher-lying transitions (Fig. 2). The 
presence of an unobserved 85-keV transition in cascade with the 265-keV line, but not with 
the 350-keV transition, was ascertained. One key evidence leading to 
the revised placements was the observation of a weak 798-keV $\gamma $ ray which is
found to be coincident with the 328-keV and all lower-lying transitions, but not with
the 533-, 350- and 265-keV ones (Fig. 3). Another crucial indicator is the observation
of a weak 334-keV transition in coincidence with all $\gamma $ rays except with the 328-keV
one (Fig. 4). The 334-keV transition is clearly visible in the delayed spectra but not 
in the prompt ones. These coincidence relationships, along with intensity balance arguments
from the delayed data, justify the revision in the relative placement of 
the transitions. Further, inspection of the prompt spectra reveals that the intensity of
the transitions reduces approaching closer to the (29/2$^{+}$) isomer, in the order 588, 350, 533,
and 328 keV, respectively. The 334- and 798-keV $\gamma $ rays may not have been identified in the 
previous work on account of their low intensity and proximity of the 798-keV $\gamma $ ray
to the strong, 7/2$^{+}$ $\rightarrow $ 3/2$^{+}$, 795-keV transition.
The newly determined levels above the $E_{x}$ = 2048-keV state are, therefore, the
ones with $E_{x}$ = 2134, 2399, 2932, 3260 and 3266 keV and {\it I}$^{\pi }$ = (19/2$^{-}$),
(19/2$^{-}$), (23/2$^{-}$), (25/2$^{+}$) and (29/2$^{+}$), respectively (Fig. 1). The 328-keV 
$\gamma $ ray deexciting the 3260-keV level is assigned $E1$ character from intensity balance 
considerations, as determined in the delayed spectra, with the 
3/2$^{+}$ $\rightarrow $ 1/2$^{+}$, 279-keV transition (using its measured conversion
coefficient) and the 5/2$^{+}$ $\rightarrow $ 1/2$^{+}$, 681-keV $\gamma $ ray. 
Similar considerations lead to 
the determination of $E2$ and $M1$ character for the 350- and 265-keV transitions, and 
rule out other multipolarities. As stated above, the presence
of an unobserved, 85-keV transition linking the 2134- and 2048-keV 
levels is inferred. For the isomeric, 334-keV $\gamma $ ray, a (29/2$^{+}$) $\rightarrow $ 
(23/2$^{-}$), $E3$ assignment is implied. The presence of an
unobserved 6-keV, (29/2$^{+}$) $\rightarrow $ (25/2$^{+}$), $E2$ transition, also 
deexciting the isomer is inferred. 
The $\gamma $ rays observed following the deexcitation of the
{\it I}$^{\pi }$ = (29/2$^{+}$) isomer are listed in Table I.

The {\it I}$^{\pi }$ = (15/2$^{-}$) state at $E_x$ = 2048 keV is determined to be
isomeric with {\it T}$_{1/2}$ = 7.9(5) ns, in agreement with the previously reported
value \cite{Slocombe}. The histogram of time difference of $\gamma $ rays below 
and above this state and a comparison with prompt transitions of similar energy 
can be found in Fig. 5. Additionally, the data indicate the possibility of
isomerism of the 2134-keV level, with a half-life up to several nanoseconds, however
the low intensity and energy of the transitions involved do not allow for
precise estimation of the value.
All $\gamma $ rays below the (29/2$^{+}$) state are found to exhibit 
delayed feeding in the 800 $\mu $s beam-off period. The integral 
variation in intensity with time of summed coincidence counts, with gates on the
232-, 265-, 279-, 328-, 533-, 588- and 795-keV $\gamma $ rays is illustrated in Fig. 6.
A value of {\it T}$_{1/2}$ = 6.6(3) $\mu $s is obtained, with an uncertainty that is 
slightly below the quoted one in the previous work, i.e. 7.7(5) $\mu $s \cite{Pfutzner}.
The measured value is associated with the half-life of the 
{\it I}$^{\pi }$ = (29/2$^{+}$) state. 

Typically, the presence of an intervening long-lived ($\sim $ $\mu $s or greater) 
isomer prevents the identification of $\gamma $ rays feeding such a level. In this
case, with the 1 $\mu $s Gammasphere coincidence window and the 6.6(3) $\mu $s
half-life, only a small number of coincidence events comprising $\gamma $ rays
below and above the isomer were recorded. By summing coincidence spectra with
gates on two strong delayed transitions below the isomer in a three-dimensional
histogram, several prompt $\gamma $ rays feeding the isomer could be identified 
[Fig. 7(a)]. The following $\gamma $ rays were found to be in coincidence with
the delayed transitions as well as with each other: 965, 211, 530, 288, 947, 438 and 
313 keV. Their ordering is based on total intensities inferred from $\gamma $-ray 
intensities, and multipolarities which, for low-energy transitions, are 
determined from intensity balance considerations after
gating on the higher-lying, 2306-keV line, combined with 
theoretical conversion coefficients \cite{Kibedi}. Detailed 
information on these transitions is presented in Fig. 1 and Table II. 
Though some random contaminant transitions are also present in Fig. 7(a), 
their origin was accounted for in terms of the deexcitation of low-lying
states from the strong reaction channels observed in the data. The spin-parity 
assignments of the levels above the (29/2$^{+}$) state are based on systematics 
\cite{Linden} as described later and should be considered tentative. Figure 7(b) is 
the prompt coincidence spectrum with gates on the 530- and 965-keV transitions.
Additionally, four other significantly weaker $\gamma $ rays, with energies 466, 
756, 1089 and 2306 keV, were also visible in prompt coincidence, and are found
to be enhanced in the summed prompt coincidence spectrum with gates on transitions 
between the (29/2$^{+}$) and (49/2$^{+}$) levels [Fig. 7(c)]. The 2306-keV
$\gamma $ ray is found to be coincident with all the above prompt
transitions and is placed above the (49/2$^{+}$) level based on intensity
considerations. It is worth noting that similar transitions with energies 2207 and 
2256 keV are observed above the four- and three-nucleon-hole states in 
$^{204}$Tl and $^{205}$Tl, respectively \cite{Broda,Wrzesinski1}. 

The presence of three isomers above the {\it I}$^{\pi }$ = (29/2$^{+}$) state
(Fig. 1) with half-lives in the range of a few nanoseconds is deduced from an inspection
of time differences of $\gamma $ rays between the (29/2$^{+}$) and (49/2$^{+}$)
levels (Fig. 8), and the following values have been obtained:
\begin{enumerate}
\item $E_{x}$ = 4442 keV, {\it I}$^{\pi }$ = (35/2$^{-}$), {\it T}$_{1/2}$ = 4.0(5) ns 
\item $E_{x}$ = 4972 keV, {\it I}$^{\pi }$ = (39/2$^{-}$), {\it T}$_{1/2}$ = 1.9(2) ns 
\item $E_{x}$ = 6957 keV, {\it I}$^{\pi }$ = (49/2$^{+}$), {\it T}$_{1/2}$ = 3.4(4) ns 
\end{enumerate}
It may be noted that the {\it I}$^{\pi }$ = (29/2$^{+}$) state in $^{203}$Tl is the
analog of the isomeric 9$^{-}$ levels in Pb isotopes \cite{Linden,Lutter,Rosengard}.

\section{Discussion}
\subsection{Isomeric states and configurations}
The high-spin isomers in Tl isotopes can be understood as resulting 
from the coupling of the {\it h}$_{11/2}$ proton hole to specific 
neutron configurations in the corresponding Pb isotones with, in particular, 
a major contribution from neutrons occupying the {\it i}$_{13/2}$ subshell.
The low-lying two-neutron-hole configurations in even-{\it A} Pb isotopes
are the $\nu $({\it i}$_{13/2}^{-1}$, {\it p}$_{1/2}^{-1}$),
$\nu $({\it i}$_{13/2}^{-1}$, {\it f}$_{5/2}^{-1}$) and
$\nu ${\it i}$_{13/2}^{-2}$ ones leading to the realization of 
isomeric {\it I}$^{\pi }$ = 7$^{-}$, 9$^{-}$ and 12$^{+}$
states \cite{Linden,Lutter}. 
The corresponding states in odd-{\it A} Tl isotones
have the configurations $\pi ${\it h}$_{11/2}^{-1}$ $\otimes $
$\nu $$^{2}$(7$^{-}$), $\pi ${\it h}$_{11/2}^{-1}$ $\otimes $
$\nu $$^{2}$(9$^{-}$), and $\pi ${\it h}$_{11/2}^{-1}$ $\otimes $
$\nu $$^{2}$(12$^{+}$) leading to {\it I}$^{\pi }$ = 25/2$^{+}$,
29/2$^{+}$ and 35/2$^{-}$ levels. The 25/2$^{+}$ and 35/2$^{-}$
states in $^{205}$Tl have been determined to be isomeric 
with {\it T}$_{1/2}$ = 2.6 $\mu $s and {\it T}$_{1/2}$ = 235 ns,
respectively \cite{Wrzesinski1}. Since the 29/2$^{+}$ $\rightarrow $
25/2$^{+}$ $E2$ transition in $^{205}$Tl has an energy of 328 keV, the 
decay of the 29/2$^{+}$ level has been found to be prompt \cite{Wrzesinski1}.
Analogous isomeric levels have not been identified in lighter, 
odd-{\it A} Tl isotopes with mass {\it A} $\le $ 199, most likely due
to the underlying weak oblate deformation and resultant collective
structures. In $^{201}$Tl, the presence of such isomeric states
has not yet been established \cite{Dasgupta}, however their existence 
cannot be ruled out. In $^{203}$Tl, under discussion here, the observed 
long-lived decay, with {\it T}$_{1/2}$ = 6.6 $\mu $s, is assigned to 
the 29/2$^{+}$ level. The excitation energies
of the 25/2$^{+}$ levels in $^{203,205}$Tl are quite similar
(3260 and 3291 keV, respectively). This may be understood as follows:
the $\pi ${\it h}$_{11/2}$ and $\nu ${\it p}$_{1/2}$ quasiparticle
energies in the two isotopes are quite similar, however the
$\nu ${\it i}$_{13/2}$ one in $^{203}$Tl is lower by virtue of its
relative proximity to the neutron Fermi level in comparison to
$^{205}$Tl. This difference is compensated to a significant extent
by the fact that the neutron pair-gap energy in $^{205}$Tl is lower
than that in $^{203}$Tl \cite{Moller}, leading to the observed similarity 
of the excitation energies of the 25/2$^{+}$ levels in the two isotopes. 
The small separation (6 keV) between the 25/2$^{+}$ and 29/2$^{+}$ levels
in $^{203}$Tl is mirrored in $^{204}$Pb, where the 7$^{-}$ and 9$^{-}$
levels are 79 keV apart \cite{Linden}. An important difference is that
the 9$^{-}$ state is lower in $^{204}$Pb in contrast to the 25/2$^{+}$
level in $^{203}$Tl. The identification of the closely-spaced pair of
levels with {\it I}$^{\pi}$ = 25/2$^{+}$ and 29/2$^{+}$ in $^{203}$Tl
thus provides insight into the difference of the magnitude of residual 
interactions of the {\it h}$_{11/2}$ proton with the respective 
two-neutron-hole configurations in $^{204}$Pb.
The transition rates for the {\it I}$^{\pi}$ = 29/2$^{+}$ isomeric decays 
are consistent with Weisskopf estimates and those in neighboring nuclei,
{\it e.g.}, $B(E3)$[334 keV] = 5.8 W.u. 
The unhindered $E3$ decay likely implies a configuration change 
from $\pi ${\it h}$_{11/2}^{-1}$ $\otimes $ $\nu $$^{2}$(9$^{-}$) 
to $\pi ${\it d}$_{5/2}^{-1}$ $\otimes $ $\nu $$^{2}$(9$^{-}$) for
the 29/2$^{+}$ $\rightarrow $ 23/2$^{-}$ transition in $^{203}$Tl.

In $^{203}$Tl, there are four neutron holes in contrast to only two in 
$^{205}$Tl. Consequently, levels with one-proton and four-neutron-hole
configurations are possible in $^{203}$Tl, unlike in $^{205}$Tl where
only two neutron holes are present. As a result, in $^{205}$Tl,
there is no intervening level between the 29/2$^{+}$ and 35/2$^{-}$
states, and the 35/2$^{-}$ state decays by a 1217-keV, $E3$ transition,
and is characterized by a half-life of {\it T}$_{1/2}$ = 235(10) ns \cite{Wrzesinski1}.
The {\it I}$^{\pi }$ = 35/2$^{-}$ state represents the maximum
spin achievable from a three-nucleon-hole configuration, resulting 
from one {\it h}$_{11/2}$ proton and two {\it i}$_{13/2}$ neutrons.
This state is found to be isomeric in $^{203}$Tl as well, with
{\it T}$_{1/2}$ = 4.0(5) ns. It decays to the 33/2$^{+}$ level
which has a five-nucleon-hole configuration, with likely mixed
character comprising 
$\pi ${\it h}$_{11/2}^{-1}$ $\otimes $ $\nu $({\it i}$_{13/2}^{-1}$, 
{\it f}$_{5/2}^{-2}$, {\it p}$_{1/2}^{-1}$) and 
$\pi ${\it h}$_{11/2}^{-1}$ $\otimes $ $\nu $({\it i}$_{13/2}^{-1}$, 
{\it f}$_{5/2}^{-1}$,  {\it p}$_{3/2}^{-1}$, {\it p}$_{1/2}^{-1}$)  
configurations. All the levels above the 35/2$^{-}$ 
state must have configurations involving five or more nucleons. 
The observed isomerism of the 39/2$^{-}$ state may be qualitatively 
understood in terms of the hindrance induced by the change of two 
neutrons in the configurations
of the initial and final states in each of these cases.
The {\it I}$^{\pi }$ = 49/2$^{+}$ state 
most likely results from the coupling of
$\pi ${\it h}$_{11/2}^{-1}$ $\otimes $ $\nu $$^{2}$(19$^{-}$),
with the neutron configuration corresponding to the relevant
level established in the $^{204}$Pb isotone \cite{Linden}.
It is noteworthy that the level structure of $^{203}$Tl above
the 29/2$^{+}$ state inferred from the present work is quite
similar to the one associated with the strongest cascade above 
the analogous 9$^{-}$ level in $^{204}$Pb (Fig. 9). 

\subsection{``Empirical" calculations}
``Empirical" calculations, based on near-neighbor systematics, 
have been performed to estimate the energies of 
isomeric states. For this purpose, the average of 1-quasineutron energies 
from neighboring, odd-{\it A} Pb isotopes 
($^{203,205}$Pb), and 1-quasiproton energies derived from the corresponding 
excited levels in $^{203}$Tl, were used \cite{ENSDF1, ENSDF2, ENSDF3}.
The neutron pair-gap energies were obtained using the five-point formula
involving odd-even mass differences \cite{Moller}. 
To calculate the energies of states with multi-nucleon configurations,
the relevant 1-quasiparticle energies were summed with the pair-gap
energy, and appropriate corrections were incorporated for the residual
interactions. The magnitude of residual interactions was determined 
from isomeric configurations in neighboring Tl and Pb isotopes and
the following values were obtained:
15/2$^{-}$ (119 keV): $\pi ${\it h}$_{11/2}^{-1}$ $\otimes $ $\nu $({\it f}$_{5/2}^{-1}$, {\it p}$_{1/2}^{-1}$);
25/2$^{+}$ (204 keV): $\pi ${\it h}$_{11/2}^{-1}$ $\otimes $ $\nu $({\it i}$_{13/2}^{-1}$, {\it p}$_{1/2}^{-1}$);
29/2$^{+}$ (160 keV): $\pi ${\it h}$_{11/2}^{-1}$ $\otimes $ $\nu $({\it i}$_{13/2}^{-1}$, {\it f}$_{5/2}^{-1}$);
35/2$^{-}$ (13 keV): $\pi ${\it h}$_{11/2}^{-1}$ $\otimes $ $\nu ${\it i}$_{13/2}^{-2}$.
The calculated energies of the above mentioned states and a comparison
with experimental values is displayed in Fig. 10. 
Sufficient experimental data are not available to determine the residual 
interactions for the five-nucleon configuration responsible for the
49/2$^{+}$ [$\pi ${\it h}$_{11/2}^{-1}$ $\otimes $ $\nu $({\it i}$_{13/2}^{-3}$, {\it f}$_{5/2}^{-1}$)]
state. Therefore, the energy for this state has been obtained by
summing the experimental value of the 29/2$^{+}$ state in $^{203}$Tl 
with that of the $\nu ${\it i}$_{13/2}^{-2}$, 12$^{+}$ level in 
$^{204}$Pb \cite{Linden}. While the empirical calculations underestimate
the experimental energies in most cases, there is fair agreement between 
the two, with the deviations being less than 300 keV for all states.  

\subsection{Shell-model calculations}
Shell-model calculations have been performed for $^{203}$Tl employing the 
KHH7B \cite{Herling} effective interaction. Since proton excitations across the 
{\it Z} = 82 shell gap and neutron ones across the {\it N} = 126 one were
not allowed, the active orbitals in the shell-model calculations 
are {\it d}$_{5/2}$, {\it h}$_{11/2}$, {\it d}$_{3/2}$ and {\it s}$_{1/2}$
for protons, and {\it i}$_{13/2}$, {\it p}$_{3/2}$, {\it f}$_{5/2}$, 
and {\it p}$_{1/2}$ for neutrons. The shell-model code Oxbash \cite{Brown} 
was used for the diagonalization of the matrices of interest. 
The deviations of the predicted energies from the experimental values
range between a few hundred keV to 0.5 MeV. The 29/2$^{+}$ state is calculated 
to be lower in energy than the 25/2$^{+}$ level, unlike what is displayed 
by the data (Fig. 10). A comparison of the excitation energies
from experiment and those from both empirical and shell-model calculations
is presented in Table III, along with the primary underlying configurations. 
The significant disagreement between the experimental and calculated values 
for the 29/2$^{+}$ and 35/2$^{-}$ states indicates the need for improving the
interactions used in the shell-model calculations.

\subsection{Octupole core excitations}
The level with $E_x$ = 9263 keV is assigned {\it I}$^{\pi }$ = (55/2$^{-}$)
and interpreted as the octupole excitation of the $^{208}$Pb core built on 
the $\pi ${\it h}$_{11/2}^{-1}$ $\otimes $ $\nu $({\it i}$_{13/2}^{-3}$, 
{\it f}$_{5/2}^{-1}$), {\it I}$^{\pi }$ = 49/2$^{+}$ configuration. The energy
of the 55/2$^{-}$ $\rightarrow $ 49/2$^{+}$ transition (2306 keV) is consistent
with this interpretation, as described below. Following the prescription outlined 
previously \cite{Wrzesinski1,Broda,Kadi}, the expected transition energy for the $E3$ 
excitation built on the five-nucleon hole, 49/2$^{+}$ state can be estimated.
While the unperturbed energy of the $E3$ excitation in $^{208}$Pb would be
2615 keV, energy shifts would result from the coupling to configurations
involving multiple nucleons. These can be expressed as the sum of energy
shifts corresponding to the individual constituents of such a configuration.
For instance, as described previously \cite{Broda,Kadi}, energy shifts of
$\Delta ${\it E} = -150 keV and +40 keV are obtained based on the 
following transitions: 17/2$^{+}$ $\rightarrow $ 11/2$^{-}$ (2465 keV) in
$^{207}$Tl, and 11/2$^{+}$ $\rightarrow $ 5/2$^{-}$ (2655 keV) in $^{207}$Pb,
respectively. These transitions represent the stretched $E3$ excitations built
on the $\pi ${\it h}$_{11/2}^{-1}$ and $\nu ${\it f}$_{5/2}^{-1}$ states in
$^{207}$Tl and $^{207}$Pb, respectively. For the  $\nu ${\it i}$_{13/2}^{-3}$
configuration, an energy shift of -262 keV has been estimated \cite{Broda} 
since the $E3$ excitation built on the $\nu ${\it i}$_{13/2}^{-3}$,
{\it I}$^{\pi }$ = 33/2$^{+}$ state has not been identified yet. Thus, for
the $\pi ${\it h}$_{11/2}^{-1}$ $\otimes $ $\nu (${\it i}$_{13/2}^{-3}$,
{\it f}$_{5/2}^{-1}$) configuration, an addition of the above mentioned
energy shifts yields a value of -372 keV, implying an estimated
55/2$^{-}$ $\rightarrow $ 49/2$^{+}$, 2243-keV transition energy. This
estimated value is lower than the observed one by only 63 keV. The situation is 
similar to that observed for the $\pi ${\it h}$_{11/2}^{-1}$ $\otimes $ 
$\nu $({\it i}$_{13/2}^{-2}$, {\it f}$_{5/2}^{-1}$), 
{\it I}$^{\pi }$ = 20$^{+}$ configuration in $^{204}$Tl \cite{Broda} 
where the estimated value is 61 keV lower than the 
experimental one. It is worth noting that even better agreement (within 
several keV) between estimated and experimental values is seen for the 22$^{+}$ state
in $^{204}$Tl \cite{Broda} and 35/2$^{-}$ level in $^{205}$Tl \cite{Wrzesinski1},
{\it e.g.}, excitations involving high-{\it j} orbitals only. This likely suggests
an underestimation of the magnitude of the repulsive interaction of 
the stretched-$E3$ excitation with the {\it f}$_{5/2}$ neutron 
hole. On the other hand, the reasonable agreement between estimated and experimental 
values appears to validate the assignment of the 2306-keV transition in $^{203}$Tl as 
the octupole core excitation built on the 45/2$^{+}$ state.

\section{Summary}
The excited level structure of $^{203}$Tl has been considerably expanded and 
is now established up to high spin with the inclusion of 25 new transitions, 
five isomeric states, and octupole core excitations built on a five-nucleon-hole 
configuration. Isomeric states with {\it I}$^{\pi }$ = (15/2$^{-}$), (35/2$^{-}$), 
(39/2$^{-}$) and (49/2$^{+}$), and respective half-lives of {\it T}$_{1/2}$ = 7.9(5) ns, 
4.0(5) ns, 1.9(2) ns, and 3.4(4) ns, have been identified.
The half-life of the previously determined long-lived decay is revised from 
7.7(5) to 6.6(3) $\mu $s, and is associated with 
the {\it I}$^{\pi }$ = (29/2$^{+}$) state. The structure of the levels fed by 
this long-lived decay is significantly modified. Gamma rays feeding the 6.6 
$\mu $s isomer have been identified through delayed-prompt coincidence measurements.
Of particular note is the observation of octupole core excitations
built on a five-nucleon hole configuration. ``Empirical" calculations, based
on near-neighbor systematics, are able to satisfactorily reproduce key states 
in the experimental level scheme, while those using the shell model appear to 
significantly underestimate the state energies in two instances. These
results contribute valuable information towards improving the understanding of
nuclear structure in the vicinity of the heaviest doubly-magic nucleus, 
$^{208}$Pb.

\section{Acknowledgments}
The authors wish to thank I. Ahmad, J.P. Greene, T.L. Khoo, A.J. Knox, 
C.J. Lister, D. Peterson, U. Shirwadkar, X. Wang and C.M. Wilson for assistance 
during the experiment. V.B. acknowledges assistance under the INSPIRE fellowship 
of the Department of Science and Technology, Government of India. 
S.K.T. would like to acknowledge support from the University Grants 
Commission, India, under the Faculty Recharge Programme. S.G.W. acknowledges 
support from the DST-INSPIRE Ph.D. Fellowship of the Department of Science 
and Technology, Government of India (Fellowship No.  IF150098).
This work is supported by the U.S. Department of Energy, Office of Science,
Office of Nuclear Physics, under award numbers DE-FG02-94ER40848 and
DE-FG02-94ER40834 (UML), DE-FG02-97ER41041 (UNC) and DE-FG02-97ER41033 (TUNL),  
and contract number DE-AC02-06CH11357 (ANL). The research described
here utilized resources of the ATLAS facility at ANL, which is a DOE Office 
of Science user facility.

\clearpage

\onecolumngrid

\begin{center}

\topcaption{
Energies and intensities of $\gamma$ rays, and excitation energies and spins of 
initial and final states in $^{203}$Tl up to the 6.6 $\mu $s
isomer. Transition energies are accurate to within 
0.5 keV. Statistical uncertainities on $\gamma $-ray intensities are listed. 
Relative intensities have been obtained from data in the delayed regime.
Transition multipolarities determined from intensity balance considerations
are noted.}

\tablehead{
\hline\hline
\multicolumn{1}{c}{E$_{\gamma}$ (keV)} &
\multicolumn{1}{c}{$E_i$ (keV)}&
\multicolumn{1}{c}{$\rightarrow $} &
\multicolumn{1}{c}{$E_f$ (keV)}&
\multicolumn{1}{c}{I$^{\pi}_i$} &
\multicolumn{1}{c}{$\rightarrow $} &
\multicolumn{1}{c}{I$^{\pi}_f$} &
\multicolumn{1}{c}{I$_{\gamma}$} &
\multicolumn{1}{c}{Multipolarity} \\
\hline}
\begin{mpxtabular}{c c c c c c c c c}
	(6)	   &$    3266 $&$ \rightarrow $&$    3260 $&$    (29/2^{+}) $&$ \rightarrow $&$    (25/2^{+}) $&     $-$  &  \\	
	(10)       &$    2048 $&$ \rightarrow $&$    2038 $&$    (15/2^{-}) $&$ \rightarrow $&$    (13/2^{-}) $&     $-$  &  \\
	(33)       &$    1218 $&$ \rightarrow $&$    1184 $&$    9/2^{+} $&$ \rightarrow $&$    7/2^{+} $&         $-$ & \\
	(85)       &$    2462 $&$ \rightarrow $&$    2376 $&$    (19/2^{-}) $&$ \rightarrow $&$    (15/2^{-}) $&      $-$   & \\
	114.9       &$    2048 $&$ \rightarrow $&$    1933 $&$    (15/2^{-}) $&$ \rightarrow $&$            - $&      $-$   & \\	
	143.5       &$   1218 $&$ \rightarrow $&$    1074 $&$    9/2^{+} $&$ \rightarrow $&$    7/2^{+} $&    $11(3)$     & \\
	232.1       &$   1450 $&$ \rightarrow $&$    1218 $&$    11/2^{-} $&$ \rightarrow $&$    9/2^{+} $&        $84(5)$ & \\
	265.0       &$   2399 $&$ \rightarrow $&$    2134 $&$    (19/2^{-}) $&$ \rightarrow $&$    (19/2^{-}) $&         $27(3)$ & {\it M}1 \\
	279.3       &$    279 $&$ \rightarrow $&$    0 $&$    3/2^{+} $&$ \rightarrow $&$    1/2^{+} $&         $84(4)$ & \\
	327.9       &$    3260 $&$ \rightarrow $&$    2932 $&$    (25/2^{+}) $&$ \rightarrow $&$    (23/2^{-}) $&         $100(5)$ &  {\it E}1 \\
	333.7	    &$    3266 $&$ \rightarrow $&$    2932 $&$    (29/2^{+}) $&$ \rightarrow $&$    (23/2^{-}) $&     $4(1)$            & \\
	350.2       &$    2399 $&$ \rightarrow $&$    2048 $&$    (19/2^{-}) $&$ \rightarrow $&$    (15/2^{-}) $&         $54(4)$ &  {\it E}2 \\
	362.1       &$    1933 $&$ \rightarrow $&$    1571 $&$             - $&$ \rightarrow $&$             -$ &         $-$ &   \\
	378.3       &$    1562 $&$ \rightarrow $&$    1184 $&$             - $&$ \rightarrow $&$       7/2^{+} $&         $-$ & \\  
	387.1       &$    1571 $&$ \rightarrow $&$    1184 $&$    - $&$ \rightarrow $&$    7/2^{+} $&         $3(1)$ & \\ 
	401.4       &$     681 $&$ \rightarrow $&$     279 $&$    5/2^{+} $&$ \rightarrow $&$    3/2^{+} $&        $10(2)$ & \\
	477.3       &$    2048 $&$ \rightarrow $&$    1571 $&$    (15/2^{-}) $&$ \rightarrow $&$    - $&         $7(2)$ & \\
	485.8       &$    2048 $&$ \rightarrow $&$    1562 $&$    (15/2^{-}) $&$ \rightarrow $&$    - $&         $4(1)$ & \\
	488.2       &$    1562 $&$ \rightarrow $&$    1074 $&$             - $&$ \rightarrow $&$  7/2^{+} $&         $-$ & \\  
	496.8       &$    1571 $&$ \rightarrow $&$    1047 $&$    - $&$ \rightarrow $&$    7/2^{+} $&         $7(2)$ & \\ 
	503.8       &$    1184 $&$ \rightarrow $&$    681 $&$    7/2^{+} $&$ \rightarrow $&$    5/2^{+} $&         $6(1)$ & \\
	533.1       &$    2932 $&$ \rightarrow $&$    2399 $&$    (23/2^{-}) $&$ \rightarrow $&$    (19/2^{-}) $&       $85(6)$  &  {\it E}2 \\ 
	537.3       &$    1218 $&$ \rightarrow $&$    681 $&$    9/2^{+} $&$ \rightarrow $&$    5/2^{+} $&         $10(2)$ & \\
	588.5       &$    2038 $&$ \rightarrow $&$    1450 $&$    (13/2^{-}) $&$ \rightarrow $&$    11/2^{-} $&         $90(6)$ & \\
	598.5       &$    2048 $&$ \rightarrow $&$    1450 $&$    (15/2^{-}) $&$ \rightarrow $&$    11/2^{-} $&         $4(1)$ & \\
	680.6       &$    681 $&$ \rightarrow $&$    0 $&$    5/2^{+} $&$ \rightarrow $&$    1/2^{+} $&         $5(1)$ & \\ 
	794.8       &$    1074 $&$ \rightarrow $&$    279 $&$    7/2^{+} $&$ \rightarrow $&$    3/2^{+} $&         $65(5)$ & \\
	797.6       &$    2932 $&$ \rightarrow $&$    2134 $&$    (23/2^{-}) $&$ \rightarrow $&$    (19/2^{-}) $&        $14(3)$ &  \\
	905.2       &$    1184 $&$ \rightarrow $&$    279 $&$    7/2^{+} $&$ \rightarrow $&$    3/2^{+} $&         $22(3)$ & \\ 
\hline
\end{mpxtabular}

\end{center}

\vskip 36pt


\begin{center}

\topcaption{
Energies and intensities of $\gamma$ rays, and excitation energies and spins of 
initial and final states in $^{203}$Tl above the 6.6 $\mu $s isomer. Transition 
energies are accurate to within 0.5 keV. Statistical uncertainities on $\gamma $-ray 
intensities are listed. Relative intensities have been obtained primarily from data in
the prompt regime. The intensities in Tables I and II should
be read independently of each other.}

\tablehead{
\hline\hline
\multicolumn{1}{c}{E$_{\gamma}$ (keV)} &
\multicolumn{1}{c}{$E_i$ (keV)}&
\multicolumn{1}{c}{$\rightarrow $} &
\multicolumn{1}{c}{$E_f$ (keV)}&
\multicolumn{1}{c}{I$^{\pi}_i$} &
\multicolumn{1}{c}{$\rightarrow $} &
\multicolumn{1}{c}{I$^{\pi}_f$} &
\multicolumn{1}{c}{I$_{\gamma}$} \\
\hline}
\begin{mpxtabular}{c c c c c c c c}
  211.0       &$    4442 $&$ \rightarrow $&$    4231 $&$    (35/2^{-}) $&$ \rightarrow $&$    (33/2^{+}) $&       $68(9)$  \\ 
  287.7       &$    5260 $&$ \rightarrow $&$    4972 $&$    (41/2^{-}) $&$ \rightarrow $&$    (39/2^{-}) $&         $32(2)$\\
  312.7       &$    6957 $&$ \rightarrow $&$    6644 $&$    (49/2^{+}) $&$ \rightarrow $&$    (45/2^{+}) $&         $17(2)$\\
  437.6       &$    6644 $&$ \rightarrow $&$    6207 $&$    (45/2^{+}) $&$ \rightarrow $&$    (43/2^{-}) $&         $26(2)$\\
  466.0       &$    9729 $&$ \rightarrow $&$    9263 $&$             - $&$ \rightarrow $&$    (55/2^{-}) $&         $-$\\ 
  530.3       &$    4972 $&$ \rightarrow $&$    4442 $&$    (39/2^{-}) $&$ \rightarrow $&$    (35/2^{-}) $&         $66(4)$\\
  755.6       &$    11107 $&$ \rightarrow $&$    10352 $&$    	     - $&$ \rightarrow $&$             - $&        $-$ \\
  947.2       &$    6207 $&$ \rightarrow $&$    5260 $&$    (43/2^{-}) $&$ \rightarrow $&$    (41/2^{-}) $&         $31(3)$\\
  965.2       &$    4231 $&$ \rightarrow $&$    3266 $&$    (33/2^{+}) $&$ \rightarrow $&$    (29/2^{+}) $&         $100(16)$\\ 
  1088.9       &$    10352 $&$ \rightarrow $&$    9263 $&$           - $&$ \rightarrow $&$    (55/2^{-}) $&         $-$\\ 
  2305.7       &$    9263 $&$ \rightarrow $&$    6957 $&$    (55/2^{-}) $&$ \rightarrow $&$    (49/2^{+}) $&         $11(3)$\\ 
\hline
\end{mpxtabular}

\end{center}

\vskip 36pt

\begin{center}

\topcaption{
Energies and spins of isomeric levels in $^{203}$Tl from experiment, ``empirical"
and shell-model calculations (see text for details).} 

\tablehead{
\hline\hline
\multicolumn{1}{c}{I$^{\pi}_i$} &
\multicolumn{1}{c}{T$_{1/2}$} &
\multicolumn{1}{c}{Configuration} &
\multicolumn{1}{c}{$E_{exp}$(keV)}&
\multicolumn{1}{c}{$E_{emp}$(keV)}&
\multicolumn{1}{c}{$E_{SM}$(keV)} \\

\hline}

\begin{mpxtabular}{c c c c c c}
$15/2^-$ & 7.9(5) ns & $\pi s^{-1}_{1/2}$ $\nu p^{-1}_{1/2}$ $i^{-1}_{13/2}$ & $2048$ & $2185$ & $1827$ \\ 
                                 
$29/2^+$ & 6.6(3) $\mu $s & $\pi h^{-1}_{11/2}$ $\nu f^{-1}_{5/2}$ $i^{-1}_{13/2}$  & $3266$ & $2999$ & $2803$ \\ 
                                 
$35/2^-$ & 4.0(5) ns & $\pi h^{-1}_{11/2}$ $\nu i^{-2}_{13/2}$ & $4442$ & $4148$ & $3899$  \\ 
 
$49/2^+$ & 3.4(4) ns & $\pi h^{-1}_{11/2}$ $\nu f^{-1}_{5/2}$ $i^{-3}_{13/2}$ & $6957$ & $6782$ & $6627$  \\

\hline 
\end{mpxtabular}
 
\end{center}

\clearpage

\begin{figure}[htb]
\centering
\includegraphics[angle=0,width=0.6\columnwidth]{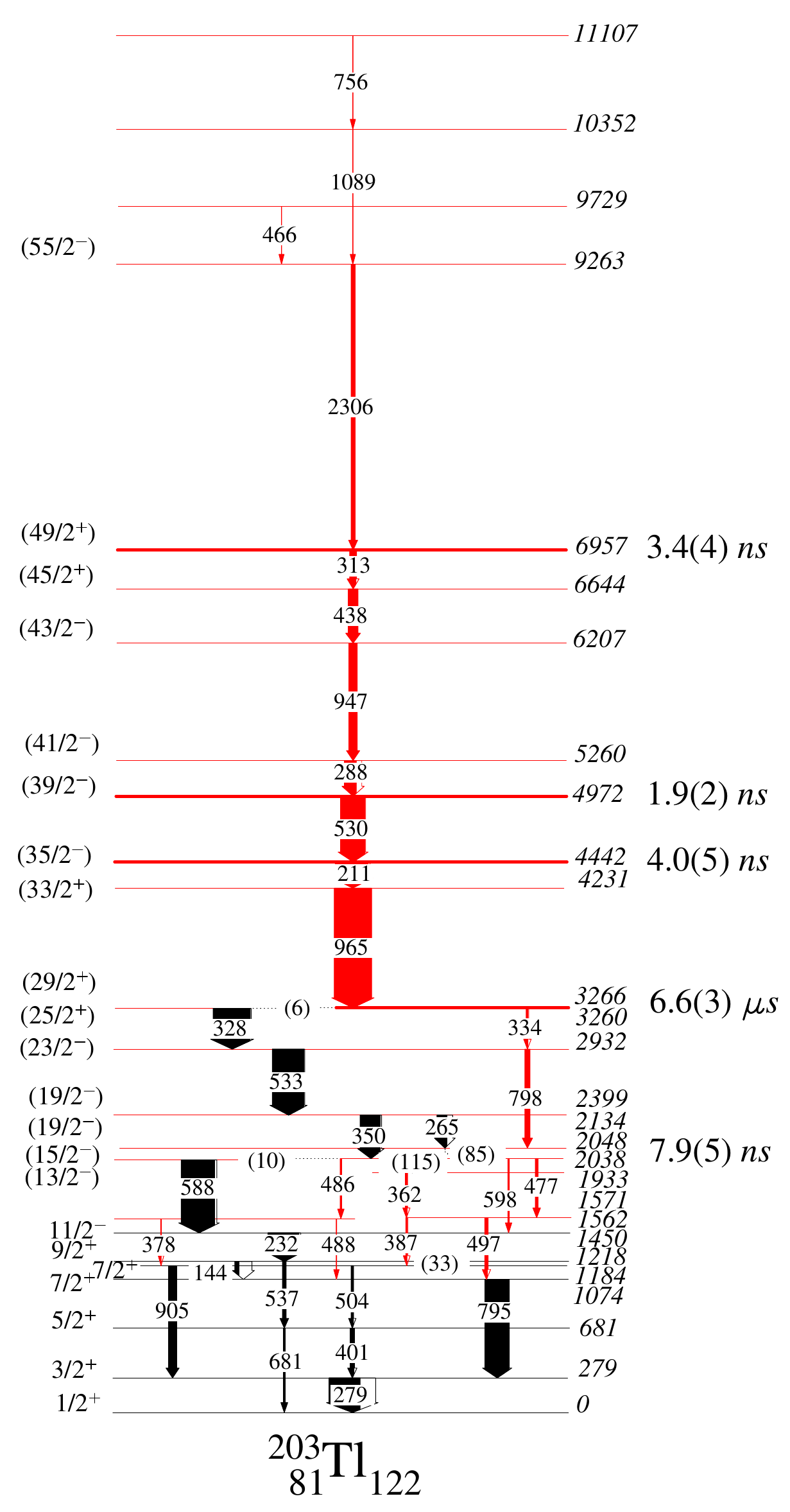}
\caption{Partial level scheme of $^{203}$Tl obtained from the present work. The transitions and levels 
above the (29/2$^{+}$) isomeric state, and several below it (marked in red), are newly established.} 
\label{fig:fig1}
\end{figure} 

\begin{figure}[htb]
\centering
\includegraphics[angle=0,width=0.6\columnwidth]{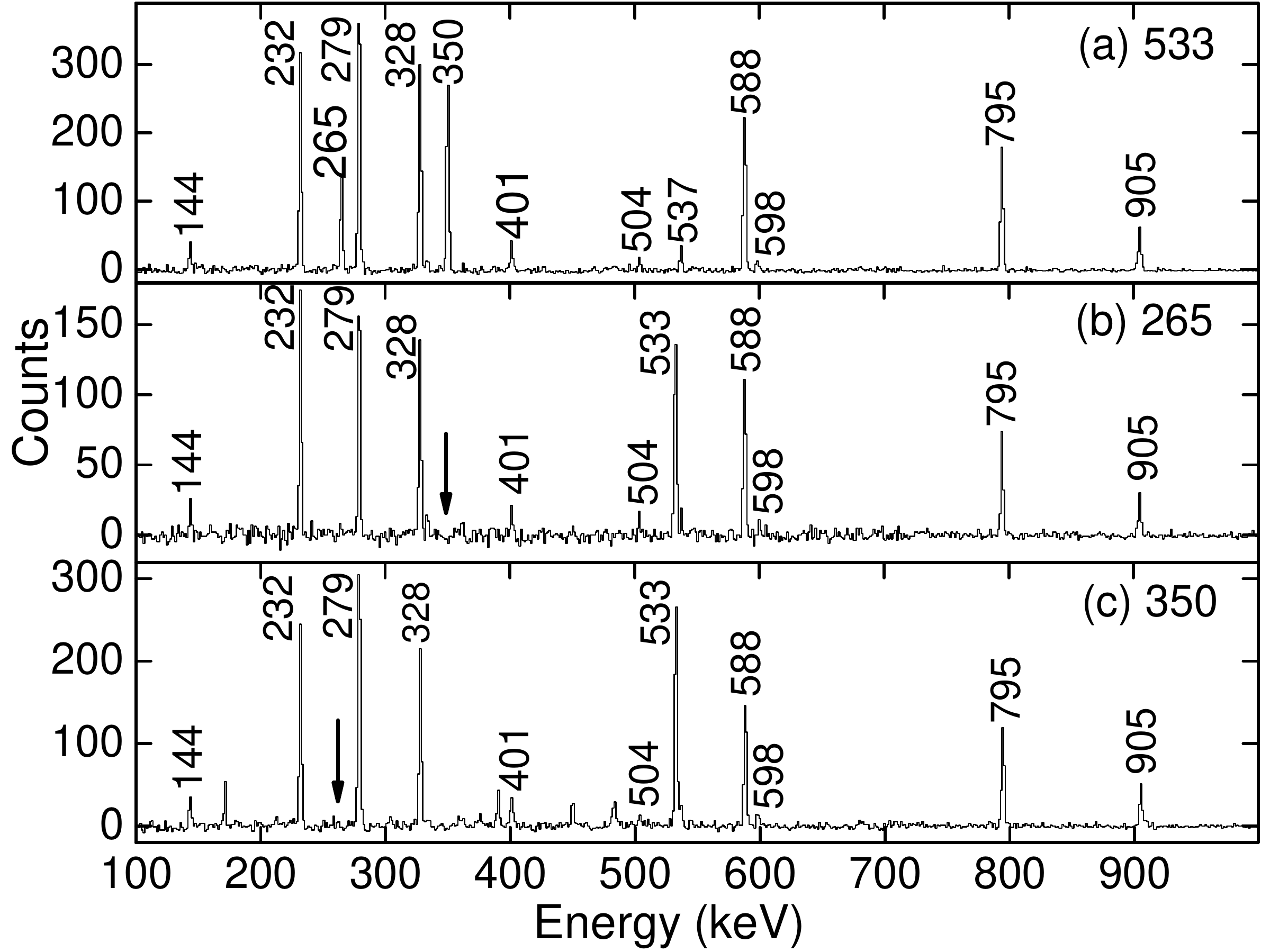}
\caption{Double-gated summed coincidence spectra showing $\gamma $ rays in coincidence with 
the: (a) 533- (b) 265- (c) 350-keV transitions, together with any of the 
232, 328 and 588-keV transitions. It is evident that the 265 and 350 keV transitions, 
the positions of which are marked with arrows in the middle and lower panels, 
are not coincident with each other as previously proposed \cite{Slocombe,Pfutzner}.}
\label{fig:fig2}
\end{figure} 

\begin{figure}[htb]
\centering
\includegraphics[angle=0,width=0.6\columnwidth]{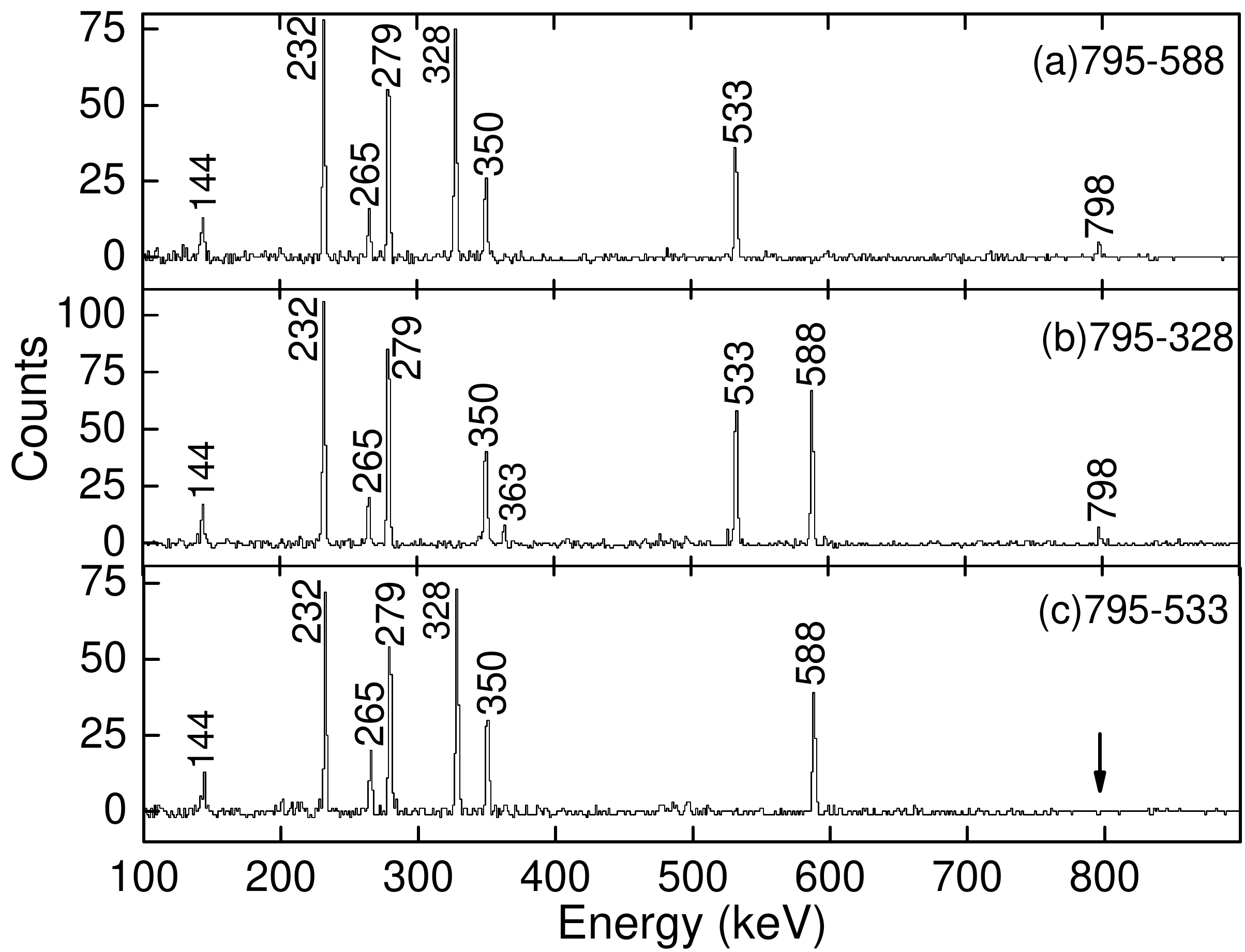}
\caption{Double-gated coincidence spectra, with gates on the transitions indicated in the 
panels, illustrating that the newly observed 798-keV $\gamma $ ray is coincident with the 
588- and 328-keV transitions, but not with the 533-keV one (arrow).}
\label{fig:fig3}
\end{figure} 

\begin{figure}[htb]
\centering
\includegraphics[angle=0,width=0.6\columnwidth]{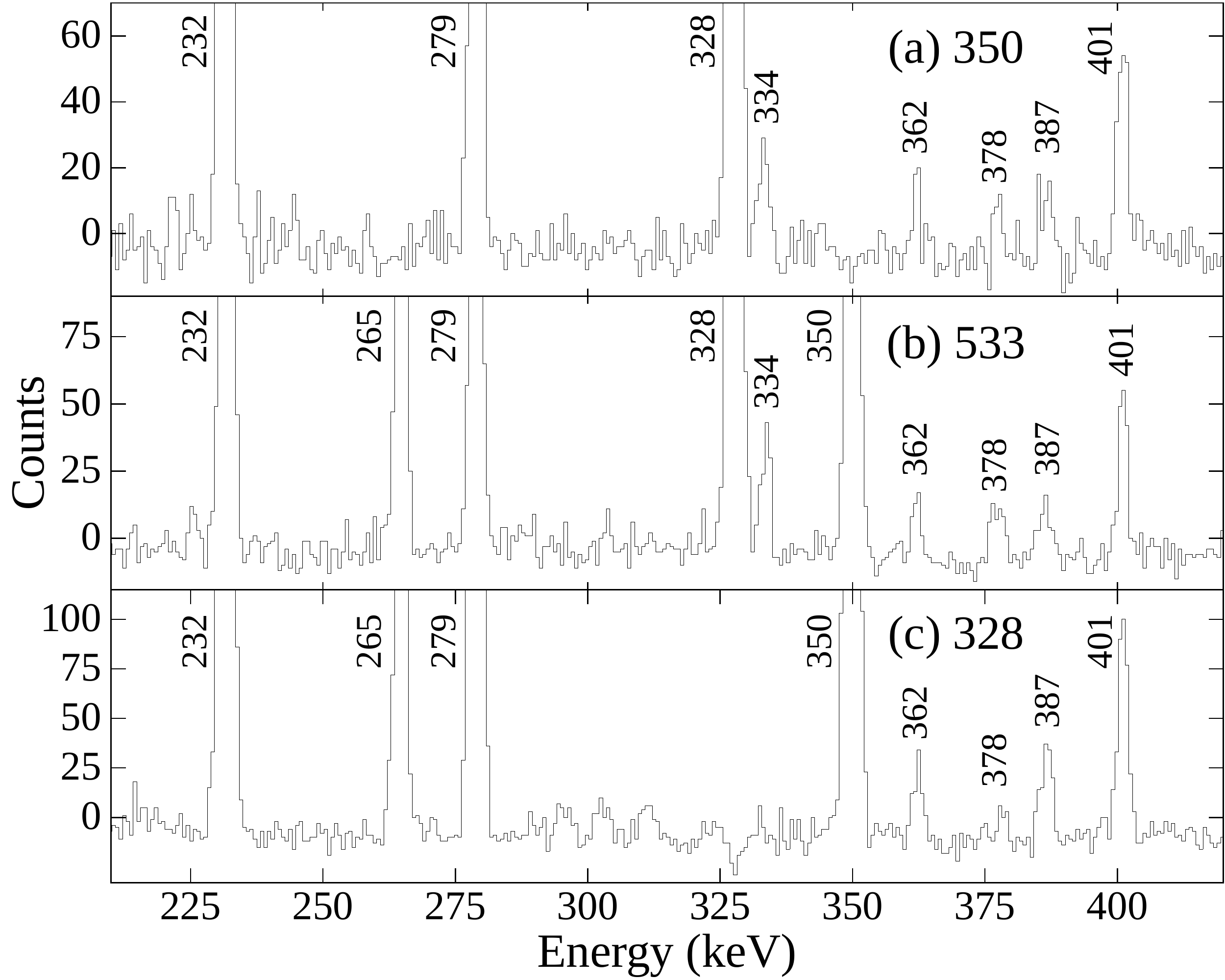}
\caption{Double-gated coincidence spectra, with gates on the $\gamma $ rays indicated in
the panels, highlighting the presence of the 334-keV isomeric transition in the 350- and
533-keV gated spectra, and its absence in the 328-keV gated spectrum.} 
\label{fig:fig4}
\end{figure} 

\begin{figure}[htb]
\centering
\includegraphics[angle=0,width=0.6\columnwidth]{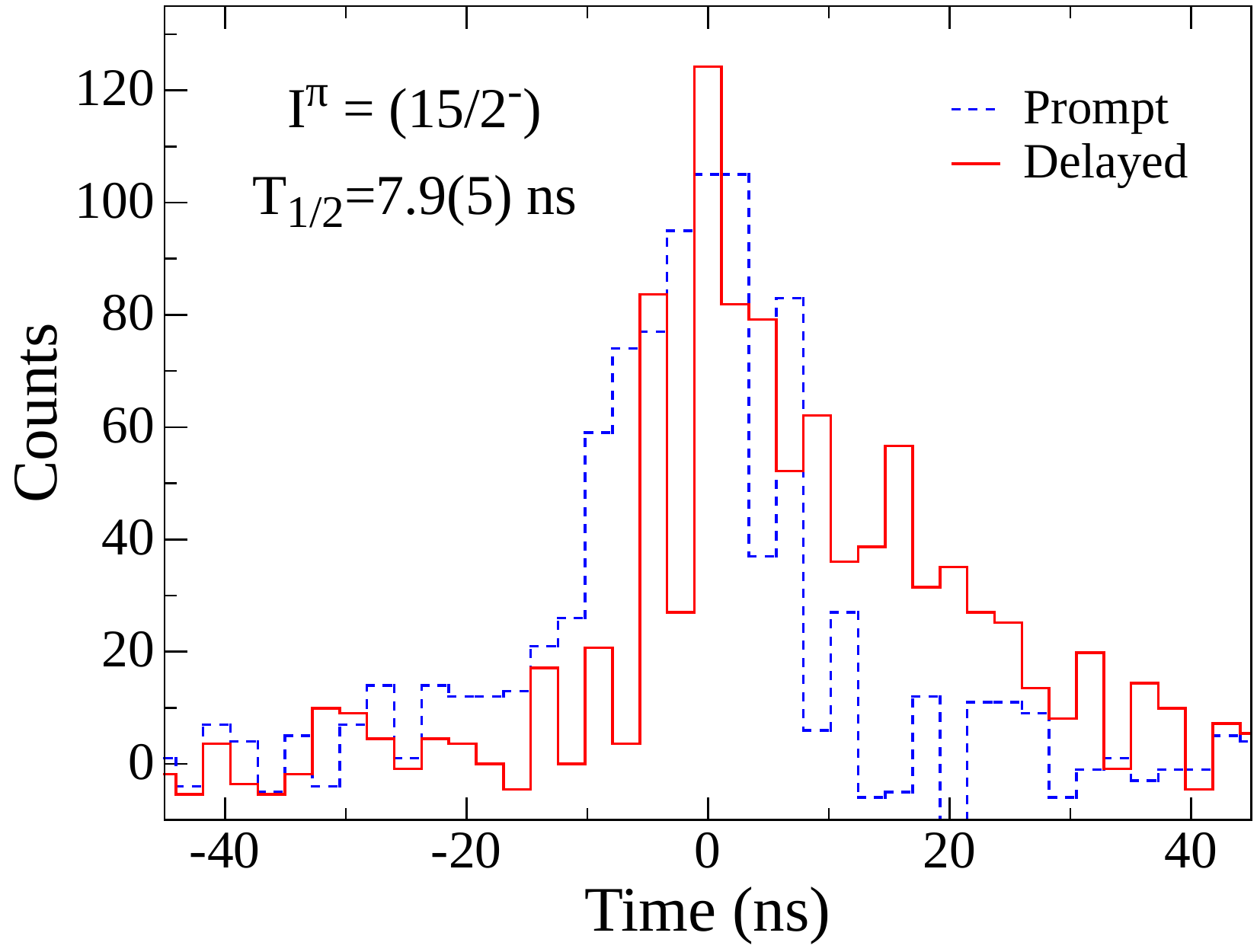}
\caption{Half-life of the {\it I}$^{\pi }$ = (15/2$^{-}$) level measured using the 
centroid-shift method. The time difference between the 588- and 350-keV transitions, below
and above this state, respectively, is displayed in red, while that of two prompt $\gamma $
rays with similar energies is shown in blue. A value of {\it T}$_{1/2}$ = 7.9(5) ns
is inferred.}
\label{fig:fig5}
\end{figure} 

\begin{figure}[htb]
\centering
\includegraphics[angle=0,width=0.6\columnwidth]{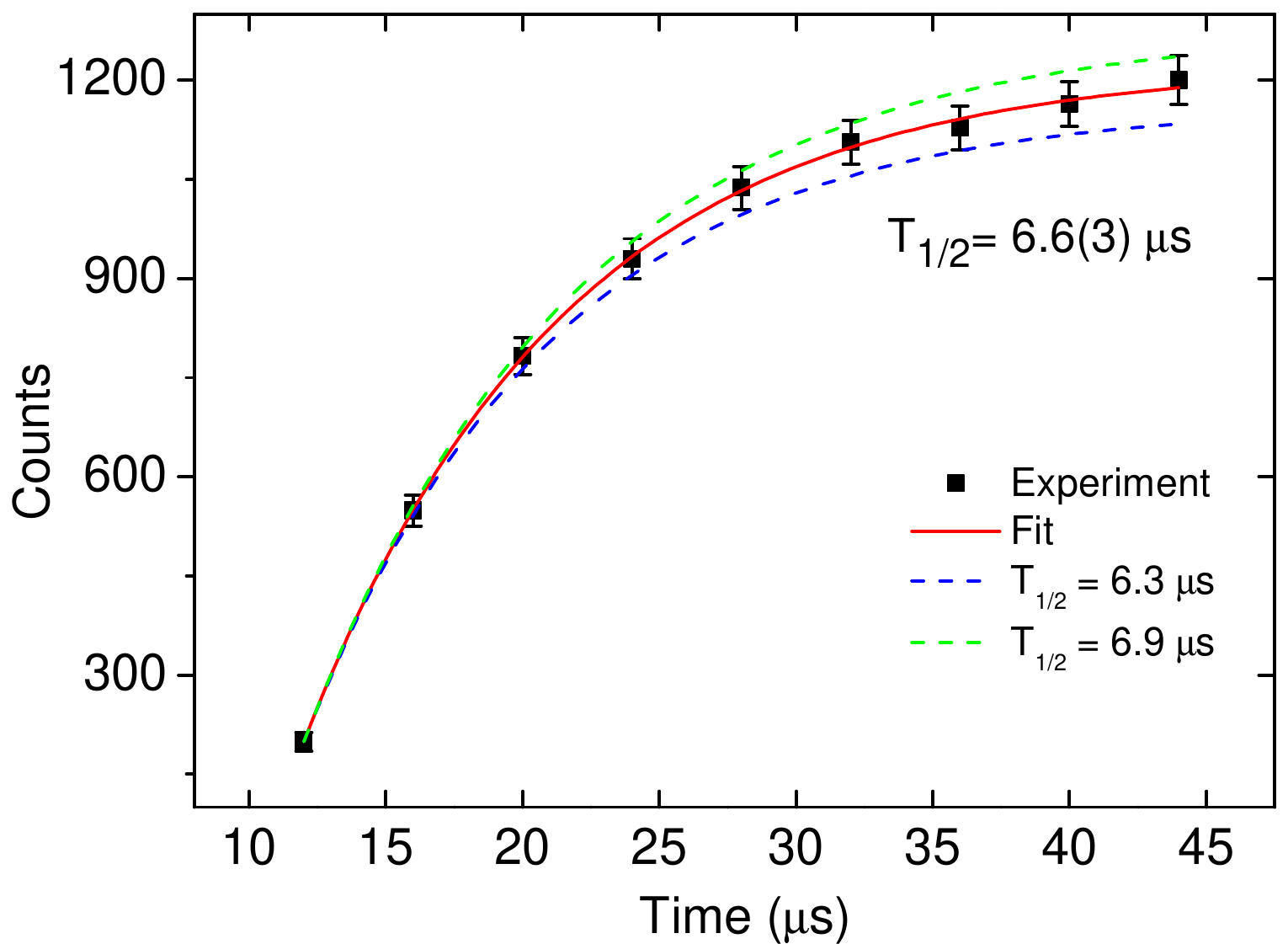}
\caption{Variation with time of the cumulative intensity
of $\gamma $ rays from the long-lived, $\mu $s decay. A half-life of
6.6(3) $\mu $s is deduced for the (29/2$^{+}$) state.} 
\label{fig:fig6}
\end{figure} 

\begin{figure}[htb]
\centering
\subfloat{%
    \includegraphics[angle=0,width=0.5\columnwidth]{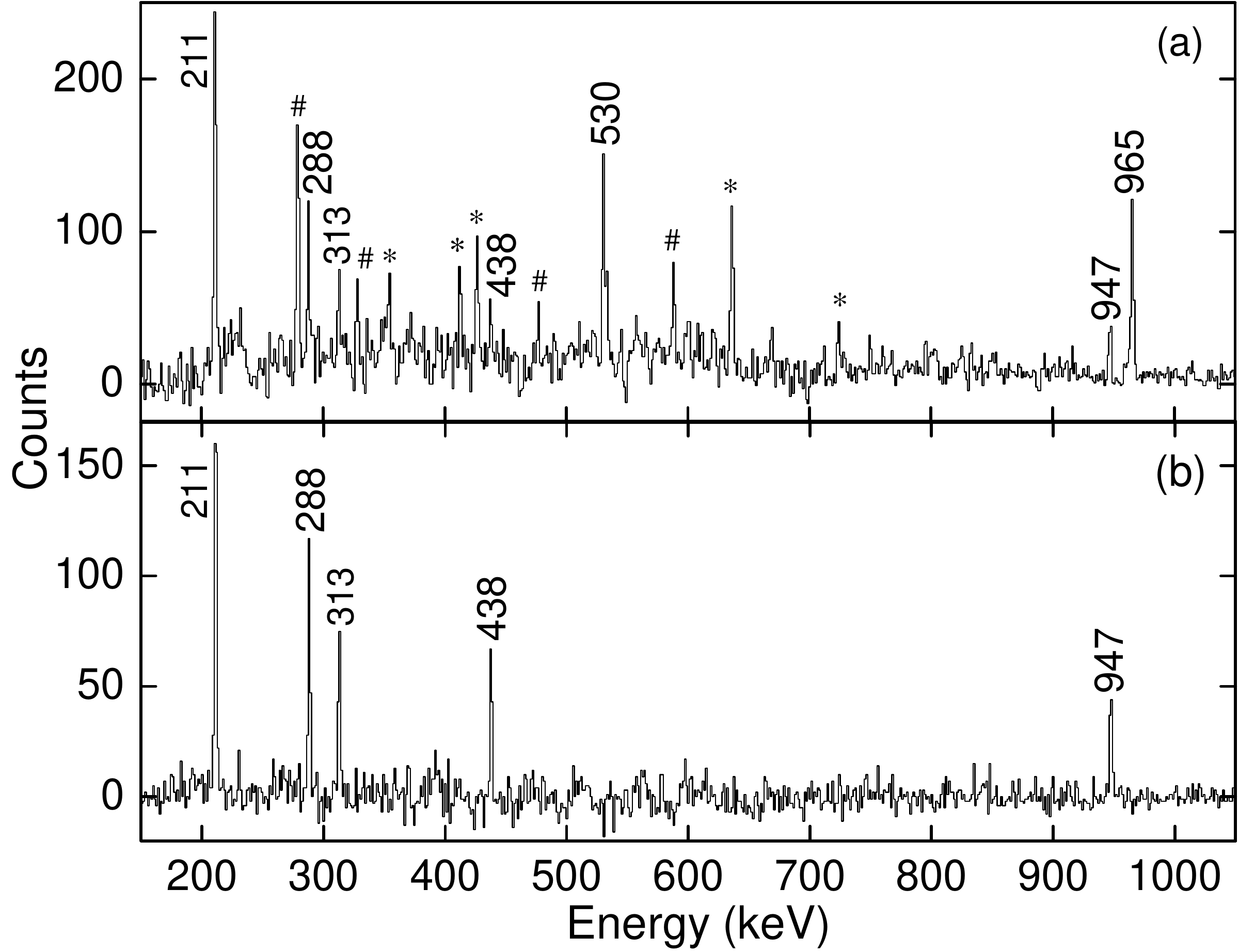}}
\subfloat{%
    \includegraphics[angle=0,width=0.5\columnwidth]{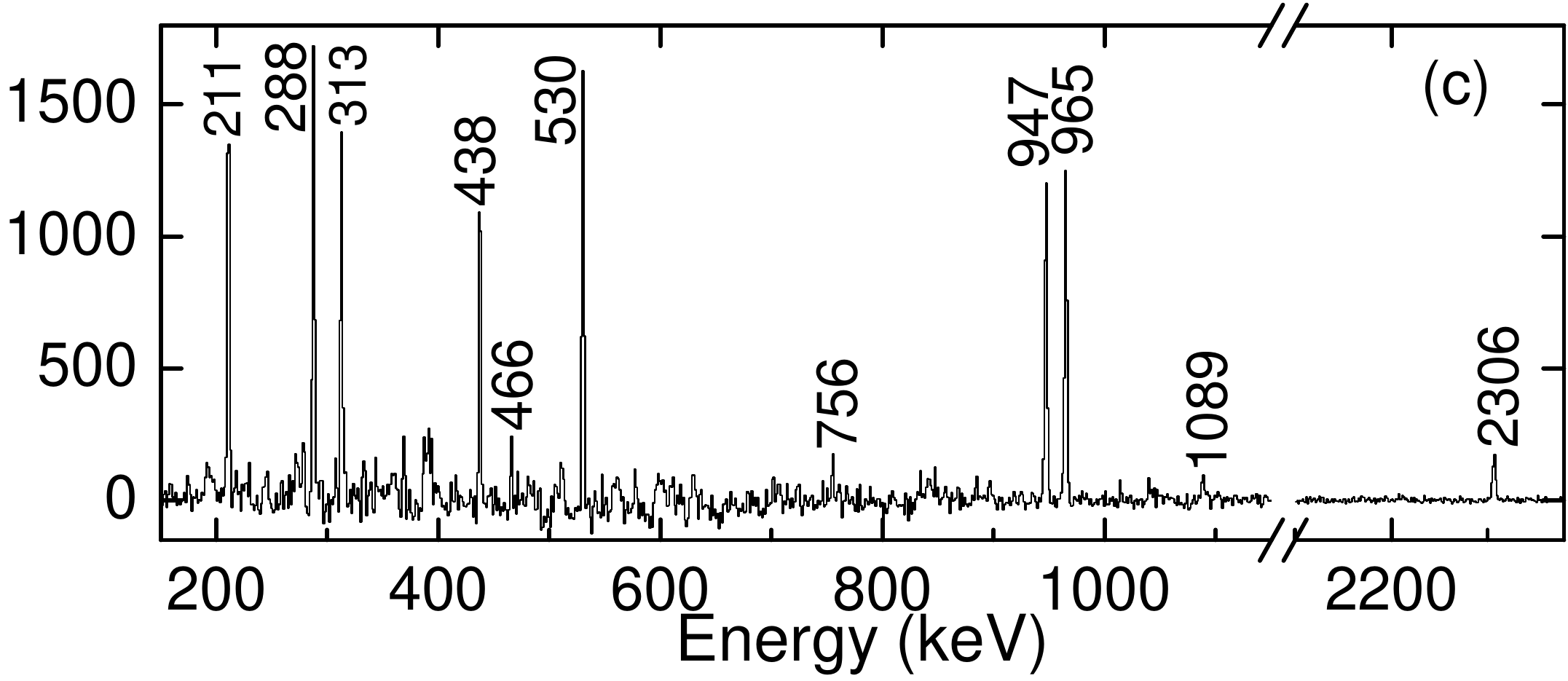}}
\caption{(a) Summed delayed-prompt coincidence spectrum with gates on the 232-, 265-,
328-, 533-, 588- and 795-keV transitions in the delayed regime. Prompt $\gamma $ rays,
feeding the $\mu $s isomer, observed in coincidence with these delayed transitions,
are displayed. The hash marks indicate strong transitions deexciting the isomer
while the asterisks denote known contaminant $\gamma $ rays from other reaction
channels. (b) Prompt coincidence spectrum with a double gate on the 530- and 965-keV
transitions. (c) Double-gated prompt coincidence spectrum obtained by summing gates
on strong transitions above the $\mu $s isomer and below the (49/2$^{+}$) level.}
\label{fig:fig7}
\end{figure} 

\begin{figure}[htb]
\centering
\includegraphics[angle=0,width=1.0\columnwidth]{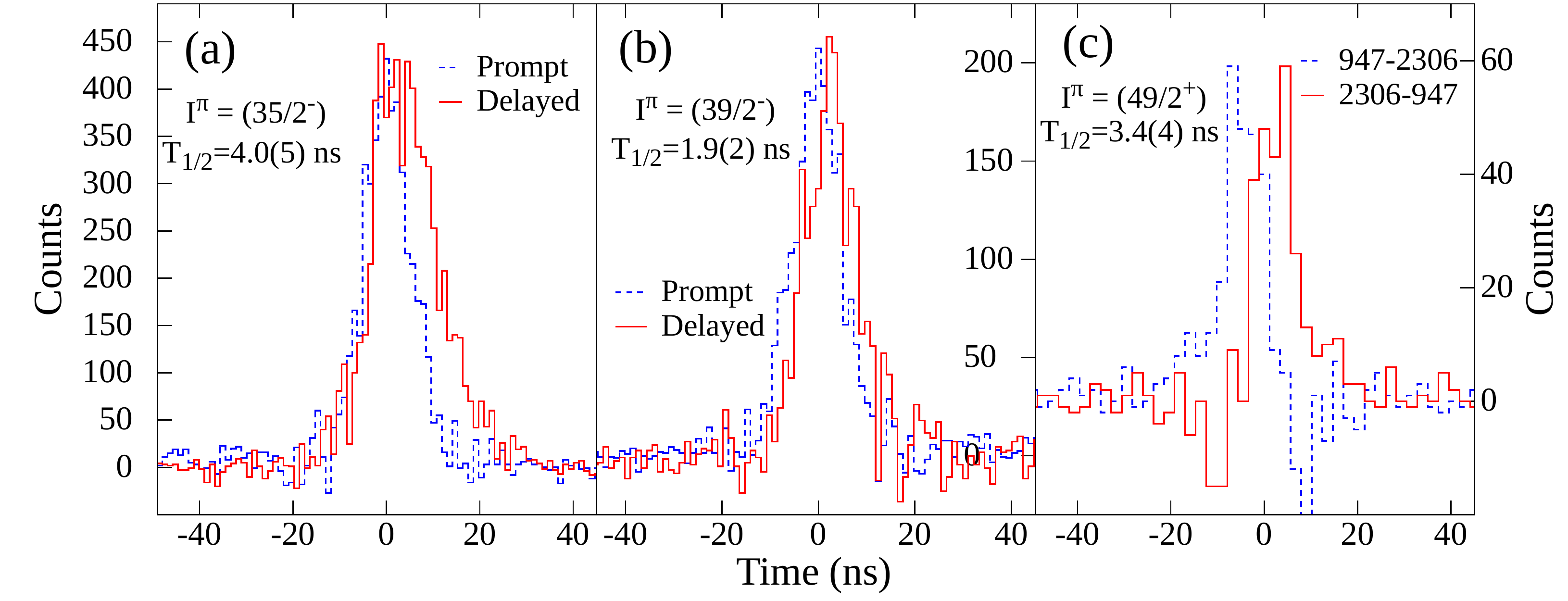}
\caption{Half-lives of the {\it I}$^{\pi }$ = (35/2$^{-}$), (39/2$^{-}$) and
(49/2$^{+}$) levels determined using the centroid-shift method. In panels 
(a) and (b), the histogram in red (solid line) indicates the time difference
of $\gamma $ rays below and above the relevant states while the one in 
blue (dashed line) corresponds to that of prompt transitions with similar 
energy. Values of {\it T}$_{1/2}$ = 4.0(5) ns and 1.9(2) ns are determined
for the (35/2$^{-}$) and (39/2$^{-}$) levels based on the observed centroid shifts.
In panel (c), the time difference between the 947- and 2306-keV $\gamma $ rays
and vice versa are compared which is equal to twice the mean life of the
{\it I}$^{\pi }$ = (49/2$^{+}$) level, leading to a value of {\it T}$_{1/2}$ = 
3.4(4) ns for this state.}
\label{fig:fig8}
\end{figure} 

\begin{figure}[htb]
\centering
\includegraphics[angle=0,width=0.4\columnwidth]{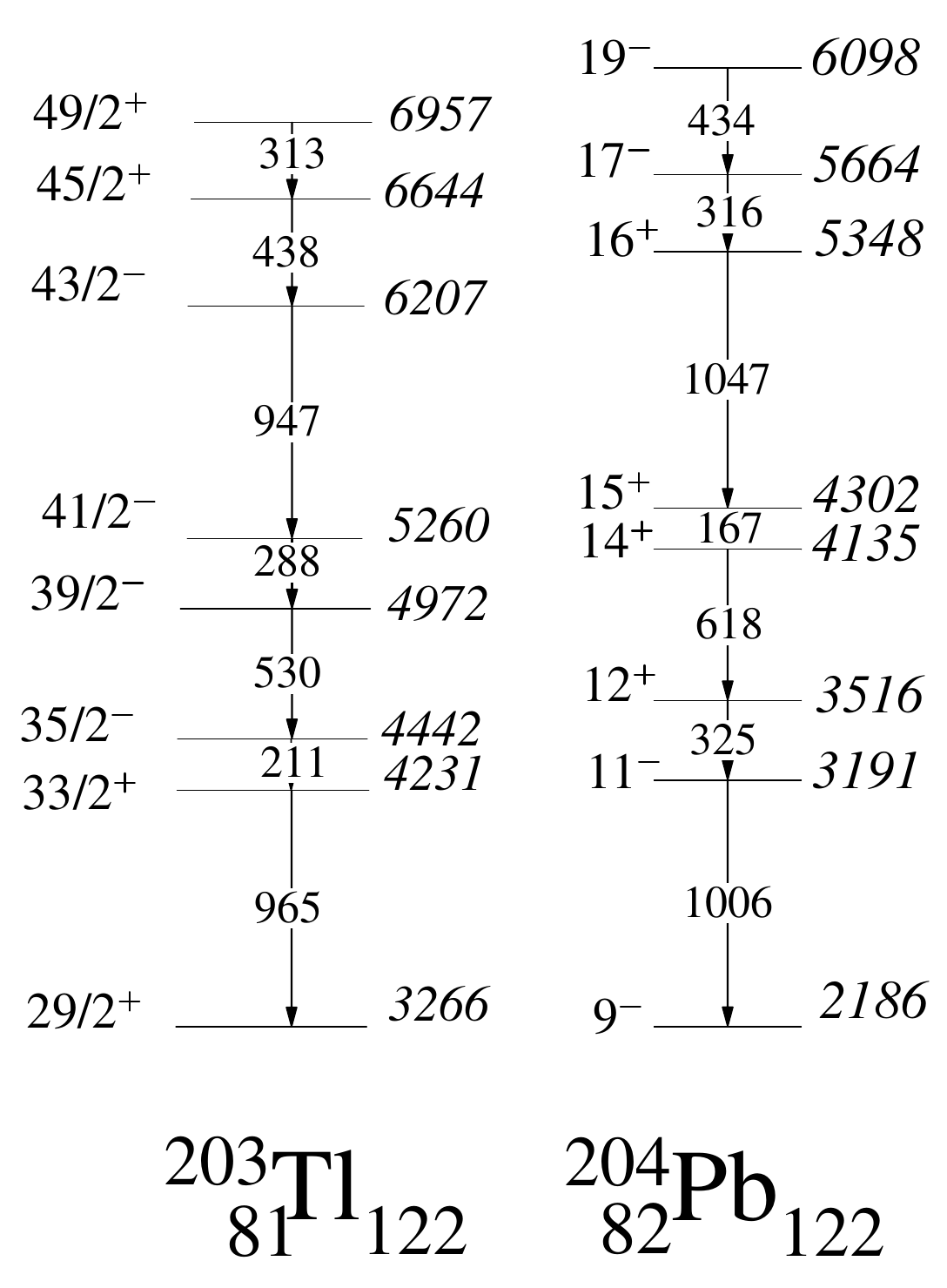}
\caption{Comparison of the level structure above the analogous 29/2$^{+}$ and 9$^{-}$
isomeric states in $^{203}$Tl and $^{204}$Pb, respectively.}
\label{fig:fig9}
\end{figure} 

\begin{figure}[htb]
\centering
\includegraphics[angle=0,width=0.4\columnwidth]{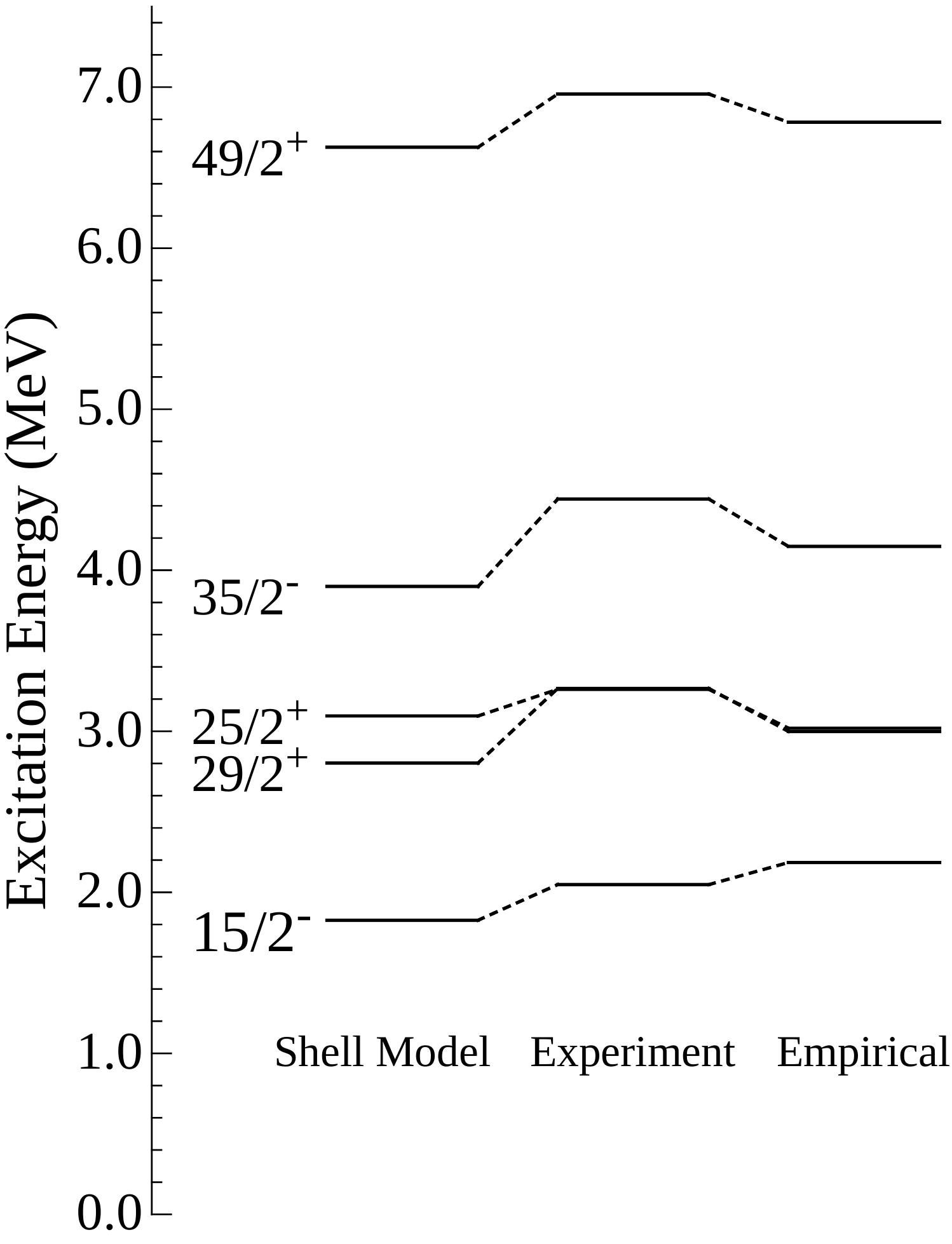}
\caption{Excitation energies of selected states in $^{203}$Tl obtained from
experiment and their comparison with empirical and shell-model calculations
discussed in the text.} 
\label{fig:fig10}
\end{figure} 

\end{document}